\documentclass[aps,prl,preprint,groupedaddress]{revtex4}
\usepackage{graphicx}

\begin{document}


\title{Diameter-dependent conductance of InAs nanowires}


\author{Marc Scheffler}
\email[]{scheffl@pi1.physik.uni-stuttgart.de}
\altaffiliation{Present address: 1.~Physikalisches Institut, Universit\"at Stuttgart, Pfaffenwaldring 57, D-70550 Stuttgart, Germany}
\affiliation{Kavli Institute of Nanoscience, Delft University of Technology, 2600 GA Delft, The Netherlands}

\author{Stevan Nadj-Perge}
\affiliation{Kavli Institute of Nanoscience, Delft University of Technology, 2600 GA Delft, The Netherlands}

\author{Leo P. Kouwenhoven}
\affiliation{Kavli Institute of Nanoscience, Delft University of Technology, 2600 GA Delft, The Netherlands}

\author{Magnus T. Borgstr\"om}
\altaffiliation{Present address: Solid State Physics, Lund University, Box 118, S-221 00 Lund, Sweden}
\affiliation{Philips Research Laboratories Eindhoven, High Tech Campus 11, 5656AE Eindhoven, The Netherlands}

\author{Erik P.A.M. Bakkers}
\affiliation{Philips Research Laboratories Eindhoven, High Tech Campus 11, 5656AE Eindhoven, The Netherlands}


\date{\today}

\begin{abstract}
Electrical conductance through InAs nanowires is relevant for electronic applications as well as for fundamental quantum experiments. Here we employ nominally undoped, slightly tapered InAs nanowires to study the diameter dependence of their conductance. Contacting multiple sections of each wire, we can study the diameter dependence within individual wires without the need to compare different nanowire batches. At room temperature we find a diameter-independent conductivity for diameters larger than 40~nm, indicative of three-dimensional diffusive transport. For smaller diameters, the resistance increases considerably, in coincidence with a strong suppression of the mobility. From an analysis of the effective charge carrier density, we find indications for a surface accumulation layer.\end{abstract}


\maketitle


Semiconducting nanowires are a focus of current research due to possible room-temperature applications in electronics\cite{Thelander2006} and optics\cite{Pauzauskie2006} as well as the possibility to study quantum phenomena in devices with an intrinsic sub-micron length scale. Amongst the variety of semiconducting materials that have successfully been grown as nanowires, InAs is of particular interest: the high mobility is advantageous for electronic applications, and quantum devices profit from the strong confinement effect. Furthermore, the simple Ohmic contacting facilitates fabrication of devices for room-temperature application as well as for low-temperature experiments.

The electronic properties of a semiconducting nanowire crucially depend on its diameter. For InAs nanowires this was noted already from the beginning of research in this field,\cite{Thelander2003,Pfund2006} but detailed studies only started very recently.\cite{Ford2009,Dayeh2009} Here one faces the experimental difficulty that studying different nanowire diameters usually requires the comparison of either different individual nanowires of one growth batch (if the nanowires have different diameters within this batch) or nanowires of different batches (if all nanowires of one batch have the same diameter). Such comparison relies on the assumption that the possibly different growth conditions for these different nanowires do not affect their electronic properties. Here we avoid these doubts by studying the diameter dependence within the \textit{same} individual nanowire: if the single-crystalline nanowire is slightly tapered and very long, different sections of this nanowire reveal the influence of the diameter.

We study nanowires of one single batch of InAs nanowires that we have grown epitaxially on an InP substrate using MOVPE and colloidal gold particles as nucleation seeds for the nanowire growth. The growth parameters are the same as for those described in Ref. \cite{VanDam2006}, except that the growth time was increased to obtain longer wires (and correspondingly a larger diameter at the nanowire base). The high quality of the nanowires used in the present study is evident from low-temperature quantum transport experiments on wires from the same batch.\cite{Scheffler2008} The nanowires have a length of typically 8~$\mu$m, and their diameter $d$ changes from less than 20~nm at the tip to more than 120~nm at the bottom end.
The nanowires were transferred mechanically onto highly doped silicon wafers covered with 285~nm of thermal oxide (which act as backgates). We used an optical microscope to locate nanowires with respect to predefined markers and designed contacts for individual wires. The contacts, consisting of 20~nm Ti and 150~nm Al layers, were deposited after electron-beam lithography and a 5~s etch in buffered HF.

\begin{figure}
\includegraphics[width=7.5cm]{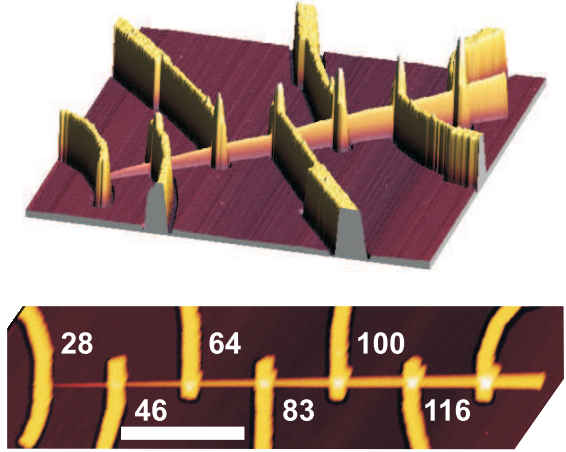}
\caption{\label{FigAFM}(Color online) AFM picture of a typical device (upper: scan range 8~$\mu$m $\times$ 8~$\mu$m, vertical axis not to scale; lower: scale bar 2~$\mu$m). The tapered nanowire is contacted with 7 metallic electrodes, each of them 200~nm wide, and separated 1000~nm from each other. The nanowire diameter, indicated in nm, is the height measured by AFM in the middle of each section.}
\end{figure}

Fig. \ref{FigAFM} shows an AFM picture of one of the devices: 200~nm wide electrodes, spaced $l = 1~\mu$m apart, define a set of nanowire sections with the same length, but varying diameter (as determined from the AFM picture and indicated in the figure). Using this arrangement of electrodes, we have measured the two-point resistance between adjacent electrodes, the corresponding backgate dependence, and the four-point resistance. The measured two-point and four-point resistances coincide, demonstrating the low contact resistance of our electrodes. From the backgate dependence we determine the mobility and charge carrier density using the established model of a conducting cylinder on top of a conducting plane.\cite{Wunnicke2006} All experiments were performed in a helium environment of 1~mbar.

\begin{figure}
\includegraphics[width=7.5cm]{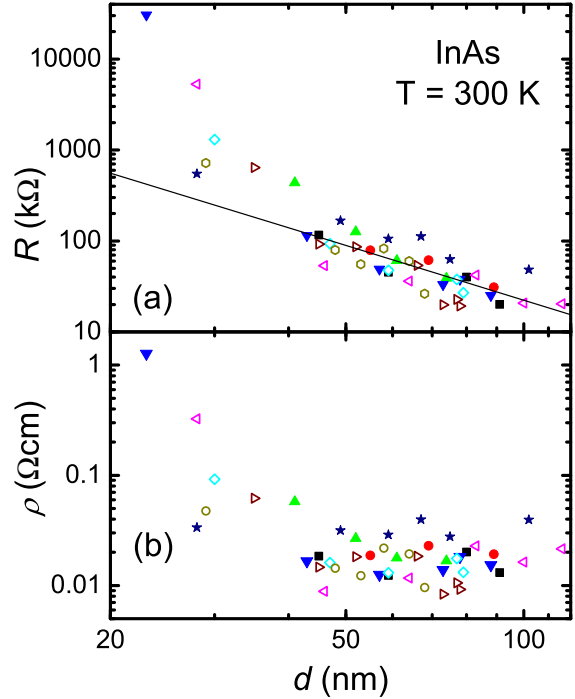}
\caption{\label{FigR2point300K}(Color online) (a) Two-point resistance of different, 1~$\mu$m long sections of tapered InAs nanowires as a function of diameter at $T =300$~K. Different symbols correspond to the nine different devices studied. (b) Resistivity determined from these data assuming three-dimensional, diffusive transport ($\rho = R \pi d^2/4l$), corresponding to the straight line in (a). This is reasonable for diameters larger than 40~nm, whereas electronic transport cannot be considered in a three-dimensional framework for smaller diameters.}
\end{figure}

First we discuss the resistance at 300~K and without applying a backgate voltage: all current-voltage and voltage-current characteristics are linear, and thus we can directly determine an Ohmic resistance. The results are shown in Fig. \ref{FigR2point300K}(a) for several nanowire devices as a function of diameter of the respective nanowire section. The resistances of all the devices follow the same overall diameter dependence. The data scattering of different devices amounts to a factor of four; this is reasonably small even when compared to recent studies of sections within the same, untapered Si nanowires.\cite{Park2008} With increasing nanowire diameter, the resistance is reduced, as expected. For diameters larger than 40~nm, the resistance decreases quadratically with the inverse diameter, as shown in Fig. \ref{FigR2point300K}(a) by the straight line. This quadratic dependence indicates three-dimensional diffusive electronic transport. Thus, for these diameters, a resistivity $\rho = R \pi d^2/4l$ (with $R$ resistance, $d$ diameter, and $l$ length of the nanowire section) can reasonably be extracted and is shown in Fig. \ref{FigR2point300K}(b). As expected from the quadratic behavior found in Fig. \ref{FigR2point300K}(a), the resistivity in Fig. \ref{FigR2point300K}(b) is basically constant for diameters larger than 40~nm. For smaller diameters, the resistance as shown in Fig. \ref{FigR2point300K}(a) increases much more strongly than following the quadratic dependence; this is equivalent to the resistivity increase in Fig. \ref{FigR2point300K}(b) for these diameters. This could be due to quantum confinement, which in InAs is expected to be effective already at rather large diameters, as soon the diameter becomes comparable to the Bohr radius (35-40~nm). Another explanation could be that the effective diameter for electronic transport in the nanowire is smaller than the physical diameter observed with AFM. Here the native oxide of the InAs nanowires, typically of the order of 2~nm thick, has to be considered, but this effect is too small to explain the strong deviations from three-dimensional behavior for diameters as big as 30~nm.

One result of particular interest that comes out of Fig. \ref{FigR2point300K} is that from these data we cannot deduce any signs of enhanced transport through a surface layer. The surface of bulk InAs is known to be conductive due to pinning of the Fermi surface within the conduction band.\cite{Mead1963} The fact that metallic contacts to InAs are typically Ohmic without Schottky barriers is also explained by such a conductive surface. Furthermore, also for InN nanowires the surface conductance can govern the electrical transport.\cite{Richter2008,Werner2009}
If such a surface layer were - within a certain diameter range - relevant for the overall transport properties of our InAs nanowires, one might naively expect that the resistance would be proportional to the inverse of the diameter, i.e. a weaker dependence than we find. But this reasoning only holds for constant electron density and mobility, whereas we will show below that both depend on diameter.
Another aspect one has to keep in mind for semiconducting nanostructures is the possibility of ballistic transport, contrasting the diffusive transport we discuss here. But as expected from studies on InAs nanowires of similar dimensions,\cite{Zhou2006} there are no signs of ballistic transport in our experiment.

\begin{figure}
\includegraphics[width=7.5cm]{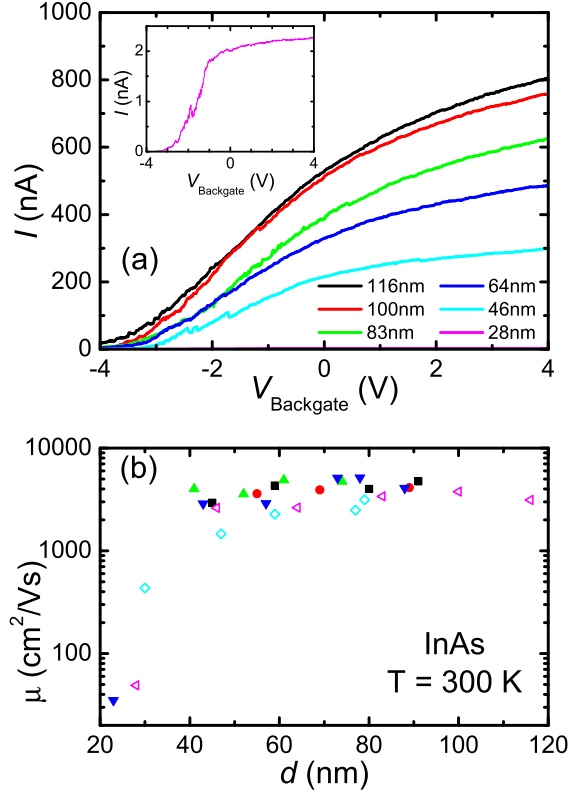}
\caption{\label{FigBGplusMob300K}(Color online) (a) Backgate sweeps for different sections within the same nanowire ($T =300$~K; same device as in Fig. \ref{FigAFM}). Applied bias voltage: 10~mV. The inset shows the data for the section with the smallest diameter. (b) Mobility determined from backgate sweeps. Different symbols correspond to the different devices studied, each with several sections of different diameter.}
\end{figure}

In backgate sweeps at a constant source-drain bias voltage $V_{\rm bias}=10$~mV, we find typical pinch-off curves as shown in Fig. \ref{FigBGplusMob300K}(a). From the slope of the linear region of these curves, we determine the mobility of the different sections of the nanowires.\cite{Wunnicke2006} We observe a pronounced diameter dependence, as shown in Fig. \ref{FigBGplusMob300K}(b): for $d > 40$~nm, the mobility is almost independent of diameter (only slightly increasing with increasing diameter), and its absolute value of 4000~cm$^2$/Vs is similar to numbers obtained previously on InAs nanowires.\cite{Bryllert2006,Ford2009,Dayeh2009} For smaller diameters, the mobility decreases dramatically.

These two effects, the increase of resistance at zero gate voltage and the decrease of the mobility, for InAs nanowires with diameter less than 40~nm are important for possible applications in room-temperature electronics: one application for InAs nanowires that is presently discussed are high-mobility field-effect transistors that might even compete with conventional Si technology.\cite{Thelander2006,Bryllert2006} However, in view of the present results, the dimensions of such a transistor might not be reduced below a channel diameter of 40~nm. This limit might be overcome by additional procedures to keep the InAs channel highly conductive, such as core-shell growth.

\begin{figure}
\includegraphics[width=7.5cm]{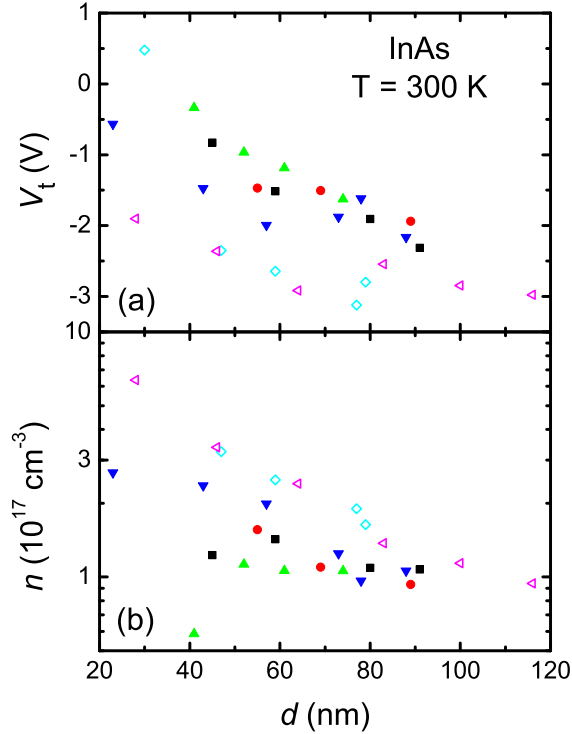}
\caption{\label{FigThresholdDensity}(Color online) (a) Threshold voltages determined from the backgate sweeps shown in Fig. \ref{FigBGplusMob300K}. (b) Electron density determined from mobility and threshold voltage. Different symbols correspond to the different devices studied, each with several sections of different diameter.}
\end{figure}

To closer investigate the transport mechanisms, we plot in Fig. \ref{FigThresholdDensity}(a) the threshold voltage $V_{\rm t}$ as determined from the linear sections of the backgate sweeps and in Fig. \ref{FigThresholdDensity}(b) the effective charge carrier density $n$ calculated via $n=C V_{\rm t}/(e \pi d^2/4 l)$, where $e$ is the electron charge and $C$ the gate capacitance.\cite{Wunnicke2006} For each nanowire device, the threshold voltage tends to become more negative for increasing diameter of the particular nanowire section.
The charge carrier density exhibits a weak diameter dependence: for decreasing nanowire diameter, the density increases. This behavior can be interpreted as due to a surface layer with increased charge carrier density. Such an accumulation layer is expected for InAs in general,\cite{Mead1963} and indications for such a diameter dependence of the density have recently been found for InAs nanowires.\cite{Dayeh2009} Interpreting our results as indications for a conductive surface layer also implies that the carrier density determined from our analysis is a mean density that averages over the nanowire cross section, without separation of bulk and surface properties.

Concerning the absolute numbers of mobility and charge carrier density, one should keep in mind that our analysis does not take into account screening of the gate field due to the metallic electrodes. This will in general lead to an overestimate of the mobility and an underestimate of the density. But this does not affect our overall results concerning the diameter dependence of mobility and density, because the characteristic screening length will play the same role for all nanowire sections studied here.

\begin{figure}
\includegraphics[width=7.5cm]{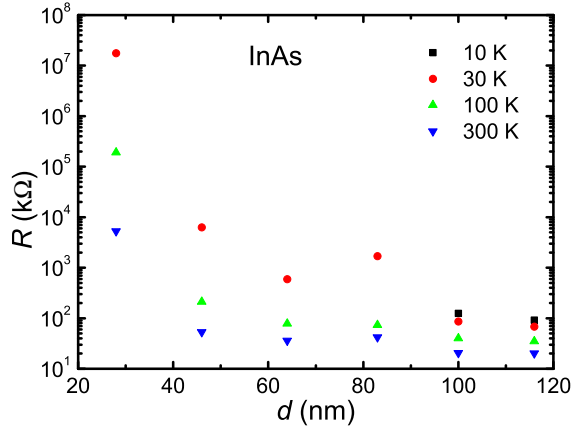}
\caption{\label{FigTemperature}(Color online) Diameter dependence for resistance at zero backgate voltage for different temperatures (same device as in Fig. \ref{FigAFM}).}
\end{figure}

Finally, in Fig. \ref{FigTemperature} we show the resistance at zero backgate for the different sections of one of the devices for a set of different temperatures. Decreasing the temperature leads to a dramatic increase of the resistance, at least for the smaller diameters. This in fact reflects a considerable shift of the depletion voltage for the backgate dependence: applying a positive voltage to the backgate can strongly increase the conductance, making low-temperature operation possible at least for the otherwise depleted devices with intermediate diameter.

In conclusion, we have shown that the influence of the diameter of a nanowire on its transport can be studied within a single nanowire if this is slightly tapered. This allowed us to establish three-dimensional diffusive transport as the relevant electronic transport mechanism at 300~K for InAs nanowires with diameters larger than 40~nm. For smaller diameters the resistance increases considerably, up to two orders of magnitude for a nanowire with 20~nm diameter. Simultaneously, the mobility decreases strongly for those small diameters, whereas it is constant for diameters larger than 40~nm. While these effects can be explained by quantum confinement, the role of a surface accumulation layer is found in the effective density of charge carriers as function of nanowire diameter.

We thank Juriaan van Tilburg and Claes Thelander for helpful discussions. This work was supported by the EU-project NODE 015783.

\end{document}